\documentclass[10pt,aps]{revtex4}
\usepackage{amssymb}
\usepackage{latexsym}
\usepackage{epsfig}
\usepackage{color}
\usepackage{float}
\usepackage{graphicx,epsfig, color}
\usepackage{graphicx}
\usepackage{subfigure}
\usepackage{hyperref}
\usepackage{amsmath}

\begin{document}
\title{\textbf{Thermal effect in hot QCD matter in strong magnetic fields}}
\author{ Xin-Jian Wen and Jia Zhang }

\affiliation{Institute of Theoretical Physics, State Key Laboratory
of Quantum Optics and Quantum Optics Devices, Shanxi University,
Taiyuan, Shanxi 030006, China}

\begin{abstract}
The quasiparticle model is improved by the free magnetic
contribution to investigated the QCD matter in a strong magnetic
field. The temperature-dependent bag function is determined by the
thermodynamic consistency to represent the difference in energy
density between physical vacuum and lowest state of QCD. It is found
that the positive bag function vanishes at high temperature
indicating the deconfinement. The rapid decrease of the bag function
in stronger magnetic fields reveals the so-called inverse magnetic
catalysis. The interaction measure at high temperature remains so
large that the usual Stefan-Boltzmann limit can not be reached. We
suggest a limit $|q_iB_m|T^2/4$ for each landau level pressure.
Finally, it is demonstrated that the positive magnetization modified
by the bag function and free magnetic contribution indicates the
paramagnetic characteristic of QCD matter.
\end{abstract}



\maketitle

\section{Introduction}
It has been observed that the strongly interacting matter,
quark-gluon plasma (QGP), produced in the Relativistic Heavy Ion
Collider (RHIC) and the Large Hadron Collider (LHC) behaves more
like a near-perfect fluid. The QCD theory in the magnetic background
may reveal a better understanding of the QGP that have rich
collective effects \cite{Lai:2000at,Koothottil:2020riy}. Many
efforts have been done in theoretical work revealing interesting
properties of strongly interacting matter in the strong magnetic
field \cite{Andersen:2014xxa,Miransky:2015ava}. With the development
of relativistic heavy ion collisions, the study of medium effect of
quark-gluon plasma becomes more active. Quark self-energy at high
temperature receives the contribution of both the electric scale and
magnetic scale, which have profound impact of confinement effects on
thermal quark collective excitation \cite{Su:2014rma,Jamal:2017dqs}.

In hot dense quark matter, one of the most important medium effects
is the effective mass generated by the nonperturbative interaction
of the particles with the system. In literature, the
phenomenological models overcome the difficulty of the QCD theory at
finite temperature and/or chemical potentials. A natural mechanism
for quark confinement is given by the MIT bag model
\cite{DeTar:1983rw}. The bag model was proposed to explain hadrons
and quark confinement, which artificially constrains the quarks
inside a finite region in space. However, the bag model is not able
to well exhibit the phase transition of the deconfinement. The quark
quasiparticle model, as an extended bag model, has been developed in
studying the bulk properties of the dense quark matter at finite
density and temperature
\cite{QPM,Thaler:2003uz,Schertler:1996tq,He:2023gva} and the
strangeles in finite size \cite{Wen:2009zza}. The quasiparticle
description has been assumed to be valid also in the case of
sufficiently high temperature
\cite{Gorenstein:1995vm,Peshier:1995ty,Levai:1997yx}. The advantage
of the quasiparticle model is the introduction of the
medium-dependent quark mass scale to reflect the nonperturbative QCD
properties \cite{Srivastava:2010xa} and color confinement mechanism
\cite{Schneider:2001nf}. The transport properties of the quark gluon
plasma have been well investigated by the noninteracting/weakly
interacting particles with effective masses \cite{Albright:2015fpa}.
The hard thermal loop(HTL) approximation can also be used to
calculate the effective quark mass, but these calculations are valid
only in the perturbative regime of QCD
\cite{Bellac:2011kqa,Peshier:2002ww,Levai:1997yx}. There are also
self-consistent quasiparticle models and single parameter
quasiparticle model
\cite{Koothottil:2018akg,Bannur:2010zz,Bannur:2007tk}. The medium
effects are taken into account by considering quarks and gluons as
quasiparticles.  Their temperature dependent masses are proportional
to the plasma frequency. More recent developments have shown that
quasiparticles with effective fugacity have been successful in
describing the lattice QCD results
\cite{Chandra:2009jjo,Chandra:2011en}, which was initially proposed
by Chandra and Ravishankar\cite{Chandra:2011en,Chandra:2007ca}.

It is well known that quarks are bound inside hadron through strong
interaction, and it has not yet been found that quarks can exist
independently. Based on this fact, the nature of quark confinement
was derived. According to lattice simulations, the deconfinement
phase transition is of first order \cite{Boyd:1996bx}. In some
phenomenological models, an order parameter of the deconfinement
transition is the Polyakov loop which is the trace of a Wilson line
along a closed loop in the time direction\cite{Li:2018ygx}. For the
SU(3) pure gauge theory, the deconfined phase transition at high
temperature corresponds to the spontaneous Z(3) symmetry breaking.
However, the study of symmetry can be complicated in QCD due to the
quark dynamics. In particular, the quark confinement and vacuum
energy density can be well described by the density-dependent bag
constant \cite{Goloviznin:1992ws,Srivastava:2010xa,Thaler:2003uz}.
In the present work, the formula of QGP has been described as a
non-interacting gas of quarks at zero chemical potential but finite
temperature, taking into account the phenomenological bag constant.
As for gluon gas, the equation of state for gluons has been
excellently described using the ideal gas approximation at high
temperatures in lattice calculations
\cite{Peshier:1994zf,Peshier:1995ty}.

The paper is organized as follows. In Section \ref{sec:model}, we
present the thermodynamics of the magnetized QGP in the
quasiparticle model. In Section \ref{sec:result}, the numerical
results for the confinement bag function and the thermodynamic
quantities in the strong magnetic field at finite temperatures. The
last section is a short summary.

\section{Thermodynamics of quasiparticle model in strong magnetic fields}\label{sec:model}

The important feature of the quasiparticle model is the medium
dependence of quark masses in describing QCD nonperturbative
properties. The quasiparticle quark mass is derived at the
zero-momentum limit of the dispersion relations from an effective
quark propagator by resuming one-loop self-energy diagrams in the
hard dense loop (HDL) approximation. The dynamical information for
gluonic degrees of freedom can also be accessed through the
effective gluon mass. In this paper, the effective quark mass $m_q$
and gluon mass $m_g$ are adopted as
\cite{Levai:1997yx,Pisarski:1989cs,Plumari:2011mk}
\begin{eqnarray}m_g(T)&=&\sqrt{\frac{1}{6}(N_c+\frac{1}{2}N_f)g^2T^2}\,
,\label{mass1}\\
m_q(T)&=&\frac{1}{2}\big( m_{i0}+\sqrt{m_{i0}^2+
\frac{N_c^2-1}{2N_c} g^2T^2}\big)\, ,\label{mass1}
\end{eqnarray}
where $m_{i0}$ denotes the quark current mass of the quark flavor
$i$. The constant $g$ is the strong interaction coupling. In order
to reflect the asymptotic freedom of QCD, one can also use a running
coupling constant $g(Q/\Lambda)$ in the equations of state of
strange matter \cite{Shirkov:1997wi}. The parameterization of the
coupling as a function of temperature close to the theory is adopted
as \cite{Peshier:1994zf}
\begin{eqnarray}g^2(T,T_c)=\frac{48\pi^2}{(11N_c-2N_f)\ln(\lambda^2(T/T_c-T_s/T_c)^2)}
(\frac{\Lambda T_c}{T})^\eta .
\end{eqnarray}
The current mass can be neglected for up and down quarks, while the
strange quark is taken to be massive. Because the vanishing current
mass is assumed for up and down quarks, Eq.~(\ref{mass1}) is reduced
to the simple form
\begin{eqnarray}m_{u,d}=\frac{1}{\sqrt{3}}g T.
\end{eqnarray}

As a typical treatment in the quasiparticle model, one implements
confinement by introducing a bag pressure, measuring the level
difference between the physical vacuum and the ground state in the
colorful world of QCD \cite{Castorina:2011ja}. To account for the
essential non-perturbative features, the corresponding
thermodynamics is based on the ideal gas partition function with
additional contributions,
\begin{eqnarray} T\ln Z=T\ln (Z_0 Z_\mathrm{vac}Z_\mathrm{mag} ),
\end{eqnarray}
where the vacuum partition depends on the bag function $T \ln
Z_\mathrm{vac}=-B(T)V$ \cite{Castorina:2011ja} and the term $
Z_\mathrm{mag}$ stands for the free magnetic field contribution. The
conventional matter contribution is introduced by $Z_0$. So the
total partition function can produce the thermodynamic quantities
from the self-consistent relation. Within the framework of the
temperature-dependent mass $m(T)$,  the pressure of the system is
expressed as
\begin{eqnarray}P(T,B_m)=\frac{T}{V}\ln Z(T,B_m)=-\sum_{i=q,g}
[\Omega_i(T,B_m)+B_i(T,B_m)]-B_0 -{\cal V}(T_0, B_m).
\end{eqnarray}
The first term $\Omega_i$ is the conventional matter contribution.
The bag constant $B_0$ stands for the vacuum energy density at zero
temperature. The variant term $B_i(T, B_m)$ resembles the
interaction term from quasiparticles and can be interpreted as the
thermal vacuum energy density. We assume Maxwell term ${\cal
V}(T_0,B_m)$ is independent on the quark mass but on the magnetic
field, which represent the free pressure of magnetic contribution
\cite{Menezes:2008qt,Mizher:2010zb,Fraga:2012fs}
\begin{eqnarray}{\cal V}(T_0,B_m)= -\sum_{i=q}\frac{N_{c}|e_iB_m|^2}{2\pi ^{2}}\left[
\zeta'(-1,x_i)-\zeta'(-1,0)
-\frac{1}{2}(x_i^2-x_i)\ln(x_i)+\frac{x_i^2}{4}\right],
\end{eqnarray}
where the parameter $x_i=m_i^2/(2|q_iB_m|)$ is defined at the
moderate temperature $T_0=150$ MeV in the thermal bag. The constant
$\zeta'(-1,0)=-0.165421...$ was introduced in Ref
\cite{Fraga:2012fs}, which is helpful to maintain a positive
magnetic pressure. The presence of ${\cal V}(T_0,B_m)$ would not
change the thermodynamically self-consistent relation $\partial
P/\partial m_i=0$ in quasiparticle model. So we have the following
differential equation
\begin{eqnarray}\frac{ d B_i}{d T}\frac{d T}{d m_i}=-
\frac{\partial \Omega_i}{\partial m_i}.\label{eq:relation}
\end{eqnarray}
So the temperature dependent term $B_i(T,B_m)$ is
\begin{eqnarray}B_i(T,B_m)=-\int_0^T \frac{\partial \Omega_i}{\partial
m_i}\frac{d m_i}{d T}d T.
\end{eqnarray}
The derivative $\partial \Omega_i/\partial m_i$ is
\begin{eqnarray}\frac{\partial \Omega_i}{\partial m_i}=\frac{d_i|e_iB_m|}{\pi^2}\sum_{\nu=0}^\infty (2-\delta_{\nu
0}) \int_0^\infty f(\varepsilon_i) \frac{m_i}{\varepsilon_i}dp_z.
\end{eqnarray}
where the fermion distribution function is
$f(\varepsilon_i)=1/({1+\exp(\frac{\varepsilon_i}{T})})$ and the
single particle energy is $\varepsilon_i=\sqrt{p_z^2+m_i^2+2\nu
e_iB_m}$ due to the quantization of orbital motion of charged
particles in the presence of a strong magnetic field along the $z$
axis \cite{Chakrabarty:1996te}. The derivative of the mass $m(T)$
with respect to the temperature is
\begin{eqnarray}
\frac{d m_i}{d T}=\frac{N_c^2-1}{4N_c}\frac{g^2
T}{\sqrt{m_0^2+(N_c^2-1)g^2T^2/(2N_c)}}+\frac{\partial m_i}{\partial
g^2}\frac{\partial g^2}{\partial T},
\end{eqnarray}
which will be simplified as $\frac{dm_i}{dT}=\frac{g}{\sqrt{3}}$ for
zero current mass and the constant coupling $\frac{\partial
g}{\partial T}=0$. If one takes into account the running coupling
$g(T)$, the derivative of the mass $m_i(T)$ with respect to the $T$
should be calculated through the full differential relation.

The entropy density, as a measure of phase space, is unaffected by
$B_i(T,B_m)$ \cite{Thaler:2003uz}, which is clearly understood from
the relation (\ref{eq:relation}). Similar to the number density, the
entropy density is written based on the fundamental thermodynamic
relation,
\begin{eqnarray}s_i=-\frac{\partial \Omega_i}{\partial T}= \frac{d_i|e_iB_m|}{\pi^2}\sum_{\nu=0}^\infty (2-\delta_{\nu
0}) \int_0^\infty f(\varepsilon_i)\frac{p^2+\varepsilon_i^2}{T
\varepsilon_i } dp_z.
\end{eqnarray}
The net effect of the bag function is to cancel the entropy density
contribution, which would arise from the dependence of the mass
$m(T)$ on the temperature. It is well known to us that the energy
density and pressure should vanish in vacuum. So the pressure should
be normalized by requiring zero pressure at zero temperature as
\begin{eqnarray}P^\mathrm{eff}(T,B_m)=P(T,B_m)-P(0,B_m).
\end{eqnarray}

The magnetization is an important thermodynamic quantity in
understanding the QCD matter \cite{Ferrer:2019xlr}. The development
of the study on the magnetization in various methods has been
summarized in Ref. \cite{Cao:2023bmk}. At zero temperature, the
magnetization is found to be positive and to be responsible for the
anisotropic pressures \cite{Felipe:2007vb,Strickland:2012vu}.  We
propose the expression of the magnetization in quasiparticle model
as,
\begin{eqnarray}{\cal{M}}=\frac{\partial P^\mathrm{eff}}{\partial B_m}=-\sum_{i=u,d,s} \left(\frac{\partial \Omega_i}{\partial
B_m}+\frac{\partial B_i}{\partial B_m}+\frac{\partial {\cal
V}_i}{\partial B_m} \right) ,
\end{eqnarray}
where the first term is the conventional contribution from the pure
quasiparticle \cite{Menezes:2015fla}. The second term demonstrates
that the effective bag function would have additional contribution
to the magnetization, which reveals the medium effect on the hot
quark matter. It can be written as,
\begin{eqnarray} \frac{\partial B_i}{\partial B_m}= \frac{N_c |e_i|m_i}{\pi^2}\sum_{\nu=0}^\infty (2-\delta_{\nu
0}) \int_0^\infty f(\varepsilon_i)\left( \frac{\nu
B_m}{\varepsilon^2}+f(\varepsilon_i)\frac{\nu
|e_i|B_m}{\varepsilon_i T}-1 \right) \frac{dp_z}{\varepsilon_i}.
\end{eqnarray}
\section{Numerical result and conclusion}\label{sec:result}

\begin{figure*}[htbp]
\includegraphics[width=2.5in]{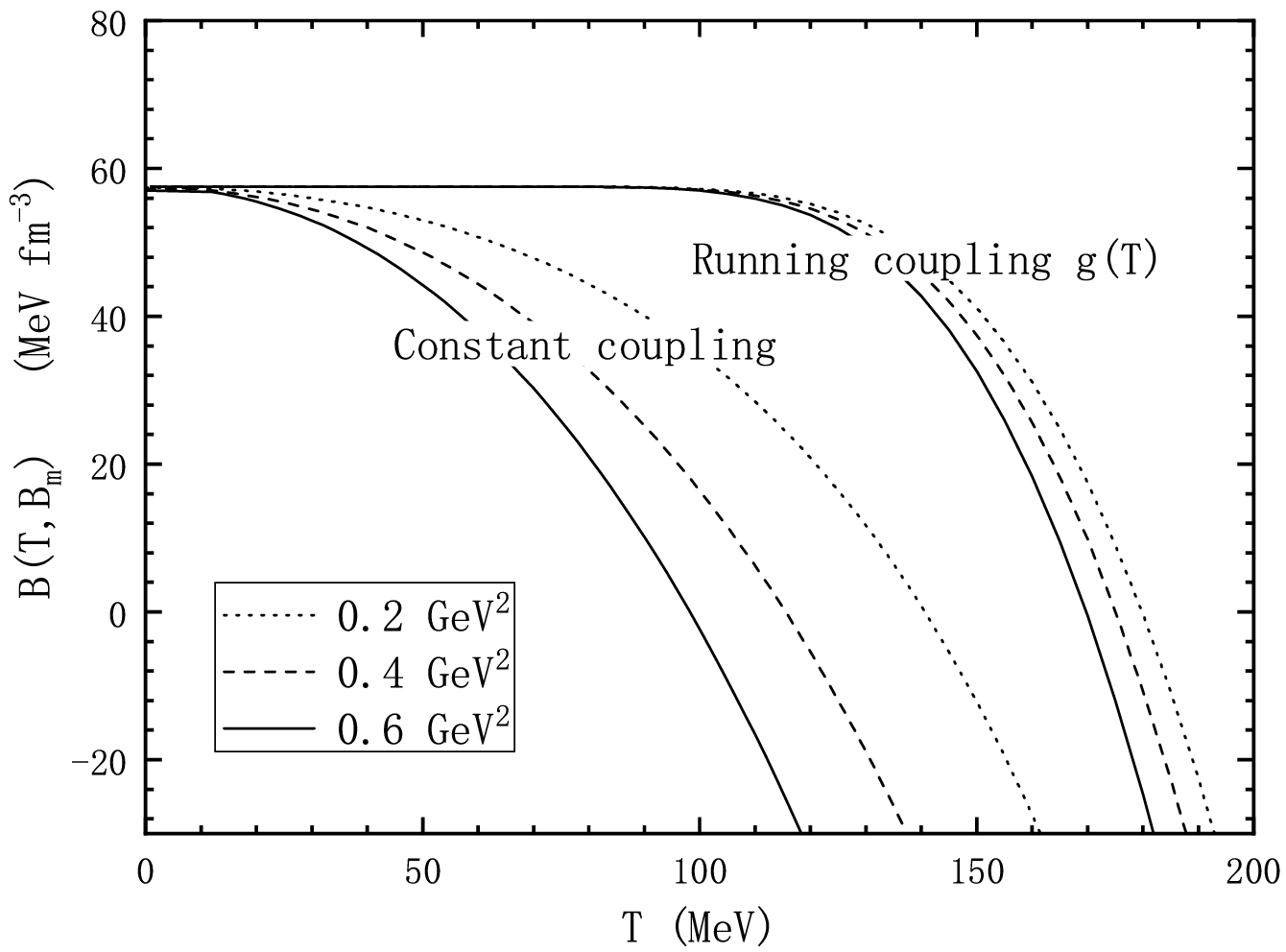}
\caption{The effective bag function $B(T, B_m)$ is shown as
functions of temperatures denoting the deconfinement transition. The
$B(T, B_m)$ decreases more rapidly with increasing temperatures at
stronger magnetic fields for both constant coupling and running
coupling constants. }\label{fig:bag}
\end{figure*}

In the framework of the preceding quasiparticle model, we have done
the numerical calculations with the quark current mass values
$m_{u}=m_d$= 0, and $m_s$=120 MeV. The constant term $B_0$ is $(145
\mathrm{MeV})^4$. The effective bag constant acts as an energy
penalty for the deconfined phase. In Fig. \ref{fig:bag}, the
effective bag constant $B(T, B_m)$ is shown as a function of the
temperature at different magnetic fields $B_m=0.2$, 0.4, 0.6
GeV$^2$. It falls to zero at high temperature, which means that the
deconfined state has larger pressure and is energetically preferred.
Compared with the fixed coupling constant $g=3$, the running
coupling constant $g(T)$ leads to a decrease of $B(T, B_m)$ at
higher temperature. The increase of the magnetic field will not
change the vacuum energy $B(T, B_m)$ at $T=0$. But the decrease of
the bag constant would be more rapidly in stronger magnetic fields,
which indicates a lower value of the critical temperature for the
deconfinement transition.

\begin{figure*}[htbp]
\includegraphics[width=2.5in]{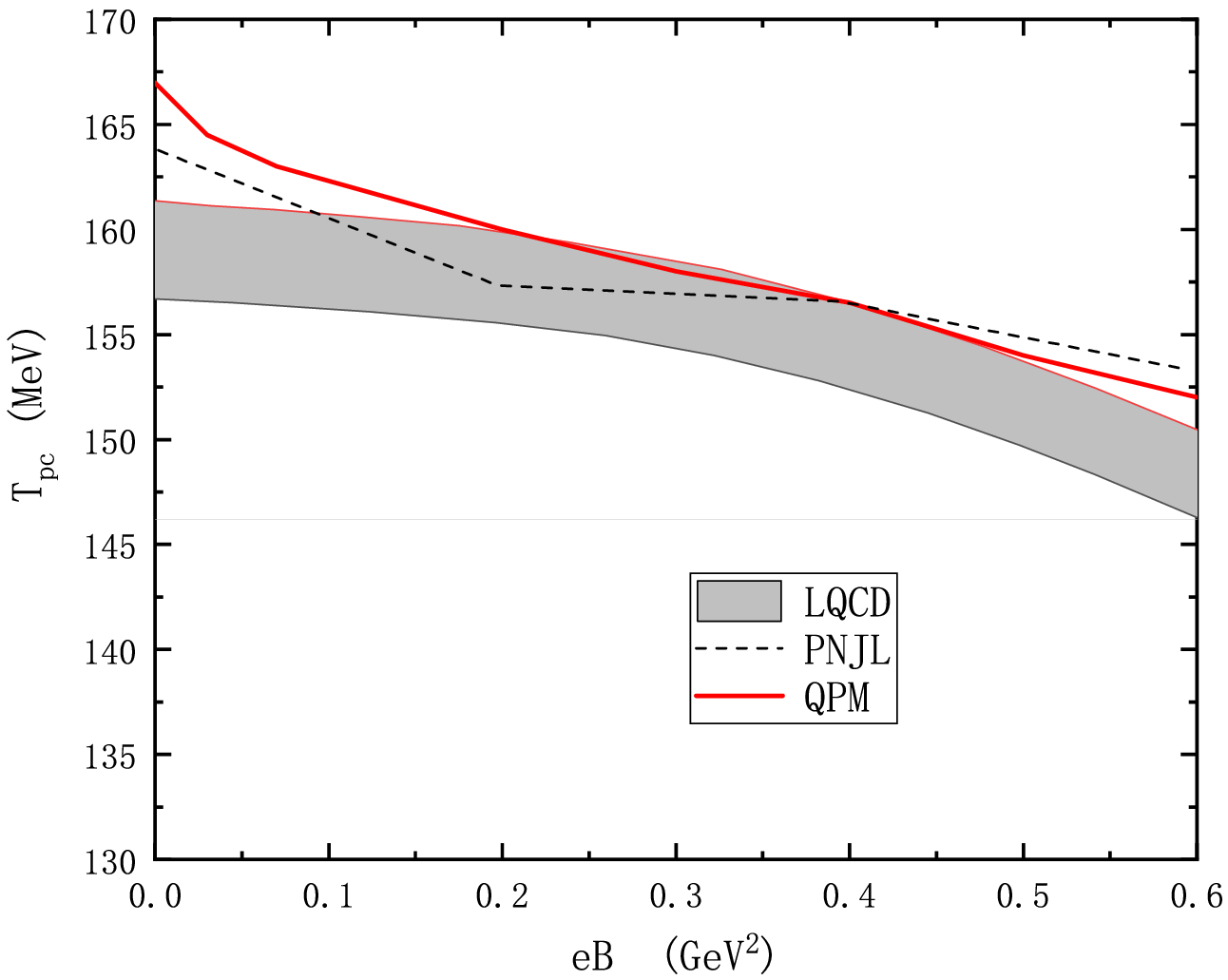} 
\caption{\label{fig:wide}The pseudocritical temperature for the
deconfinement transition in quasiparticle model (red curve) compared
with the results of the PNJL model (dashed curve)
\cite{Tavares:2021fik} and of LQCD (shadow band )\cite{Bali:2011qj}.
 }\label{fig:T}
\end{figure*}

In the quasiparticle model, the decrease of the effective bag
constant $B(T,B_m)$ denotes the deconfinement transition. The
critical temperature can be determined by the position of
half-height of the bag constant. In Fig. \ref{fig:T}, the
pseudocritical temperature is plotted by the red solid curve. For
the convenience of comparison, the result from PNJL is marked by the
black dashed curve \cite{Tavares:2021fik}. It can be clearly seen
that the trend of the decrease of the $T_{pc}$ with the magnetic
field is close to the LQCD result marked by the shadow band in the
panel \cite{Bali:2011qj}. Moreover, the pseudocritical temperature
decreases by about 10 percent of its original value at zero magnetic
field. Our result is in agreement with the so-called inverse
magnetic catalysis effect revealed by the LQCD.

\begin{figure*}[htbp]
\includegraphics[width=2.5in]{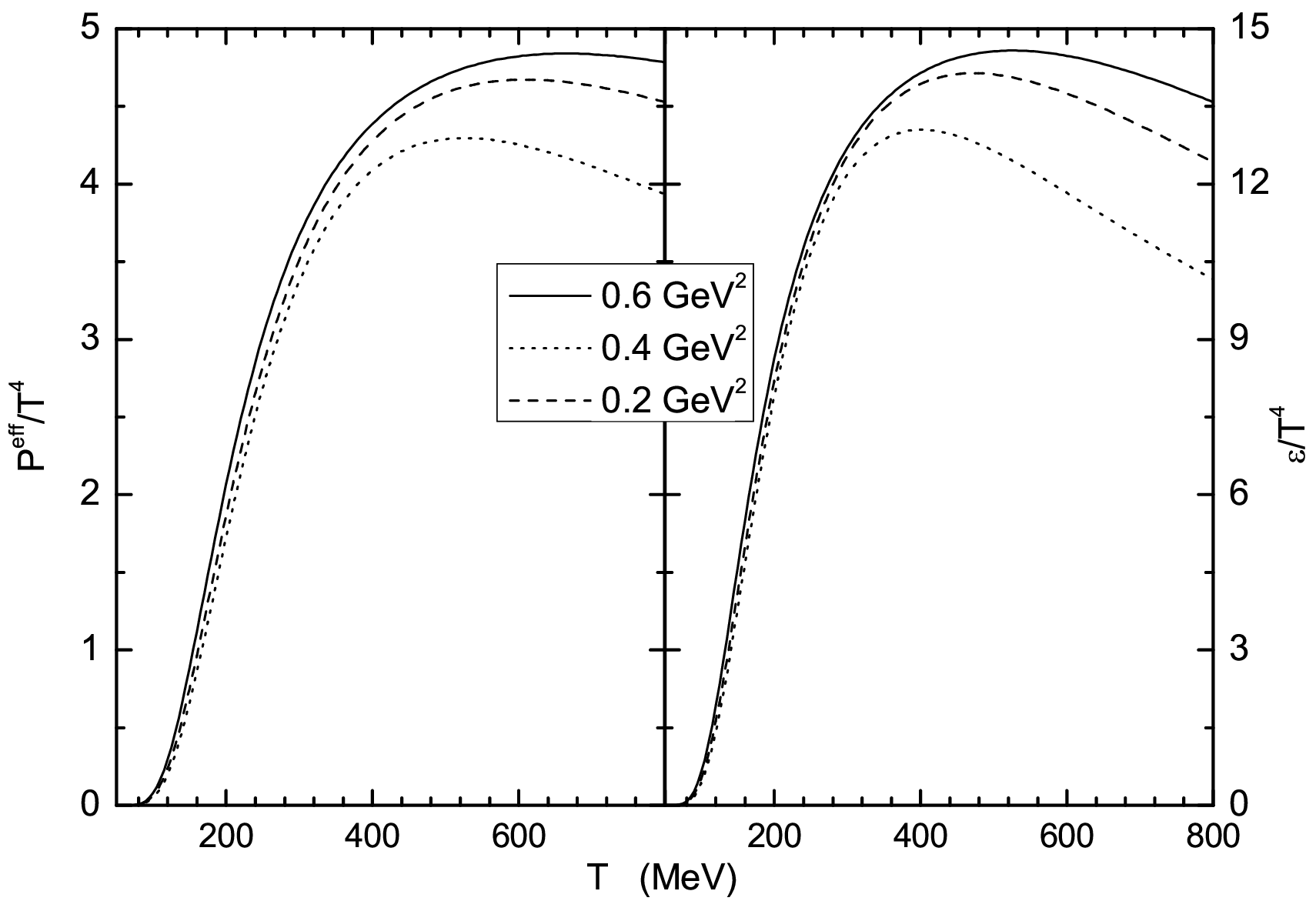} 
\caption{\label{fig:wide}The scaled pressure $P^\mathrm{eff}/T^4$ on
left panel and energy density $\epsilon/T^4$ on right panel are
always increasing functions of the temperature. It is apparent that
the Stefan-Boltzmann (S-B) limit is absent at high temperatures in
strong magnetic fields. }\label{fig:press}
\end{figure*}

Fig. \ref{fig:press} illustrates how the pressure and the energy
density (in unit of $T^4$) depend on the temperature of the medium.
In the region around and just above the critical temperature, the
energy density rises much more rapidly than the pressure, which
gives rise to a observed rapid increase of the interaction measure.
At stronger magnetic fields, the larger pressure and energy density
are obtained at the high temperature. However, the temperature is
not the only scale with the dimension of the energy. The pressure as
well as the energy density can not asymptotically converge to their
Stefan-Boltzmann value $P^\mathrm{SB}/T^4=$constant at $T\rightarrow
\infty$. The deviations from the usual Stefan-Boltzmann values are
due to the quarks are constrained by the landau level in strong
magnetic fields.

\begin{figure*}[htbp]
\includegraphics[width=2.5in]{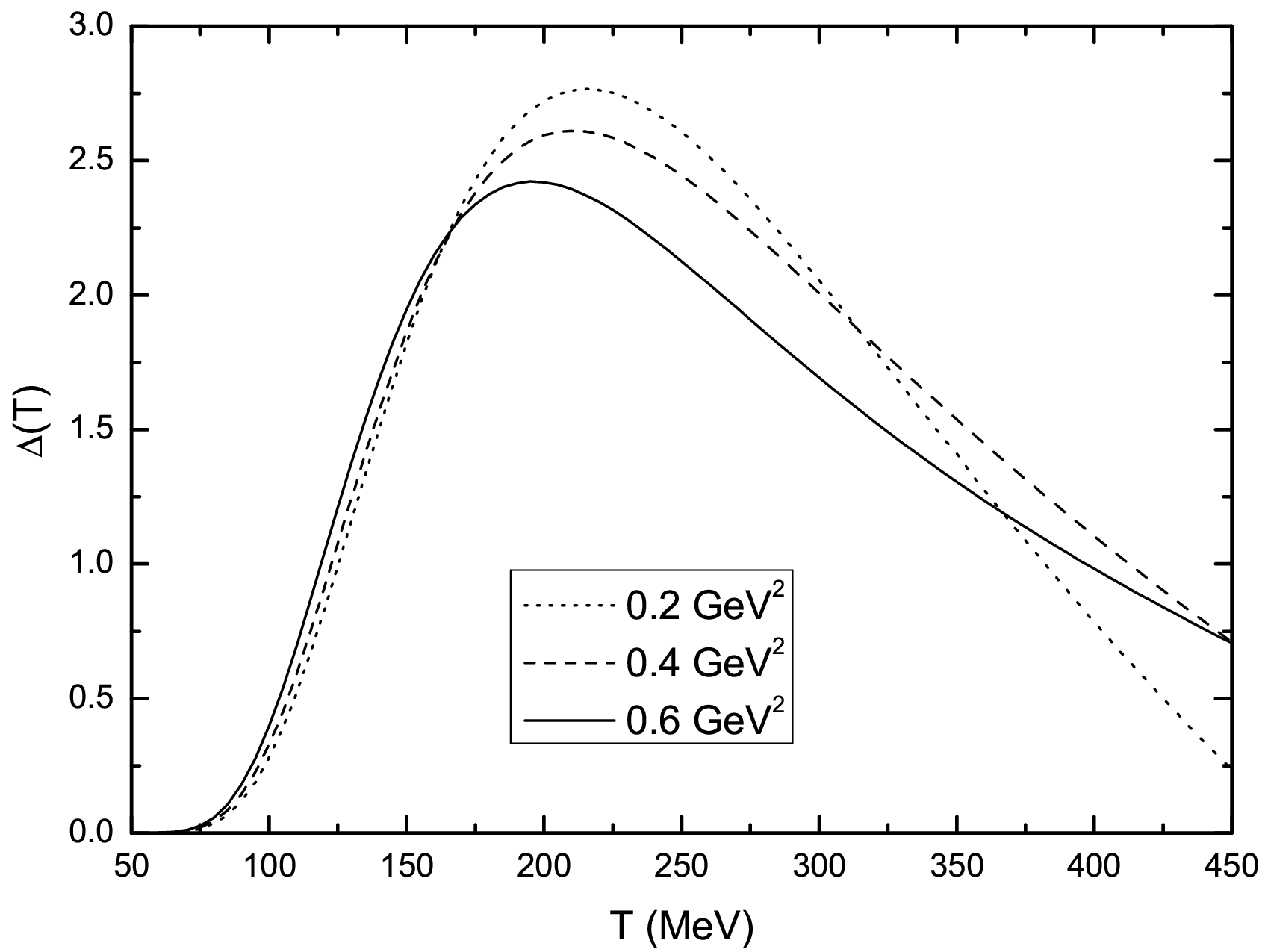} 
\caption{\label{fig:wide}The interaction measure is shown as
functions of the temperatures in different magnetic fields. It
remains so larger at high temperatures that the interaction must be
still present at $eB_m=0.6$ GeV$^2$ marked by the solid curve.
}\label{fig:measure}
\end{figure*}

Interaction measure is the trace of energy-momentum tensor. For
non-interacting massless constituents (the ``conformal" limit), is
zero, so that the temperature is the only scale. In a strong
magnetic field, the interaction measure is defined as
$\Delta(T)\equiv (\varepsilon-3 P)/T^4$ for quark gluon plasma. In
Fig. \ref{fig:measure}, the trace anomaly of the energy momentum
tensor is plotted as a function of temperature. The so-called
interaction measure normalized by $T^4$ gives the deviation from the
free gas relation between the energy density and the pressure and is
also a measure of the breakdown of conformal symmetry. Even though
the temperature dependent coupling constant is employed to realize
the asymptotic freedom at high temperature, the non-zero value
indicates that some interactions must still be present due to the
landau levels in magnetic fields. The $\Delta(T)$ at high
temperature will remain larger at stronger magnetic fields, which is
in agreement with LQCD that the interaction measure remains large
even at very high temperatures, where the Stefan-Boltzmann (S-B)
limit is not yet reached.

\begin{figure*}[htbp]
\includegraphics[width=2.5in]{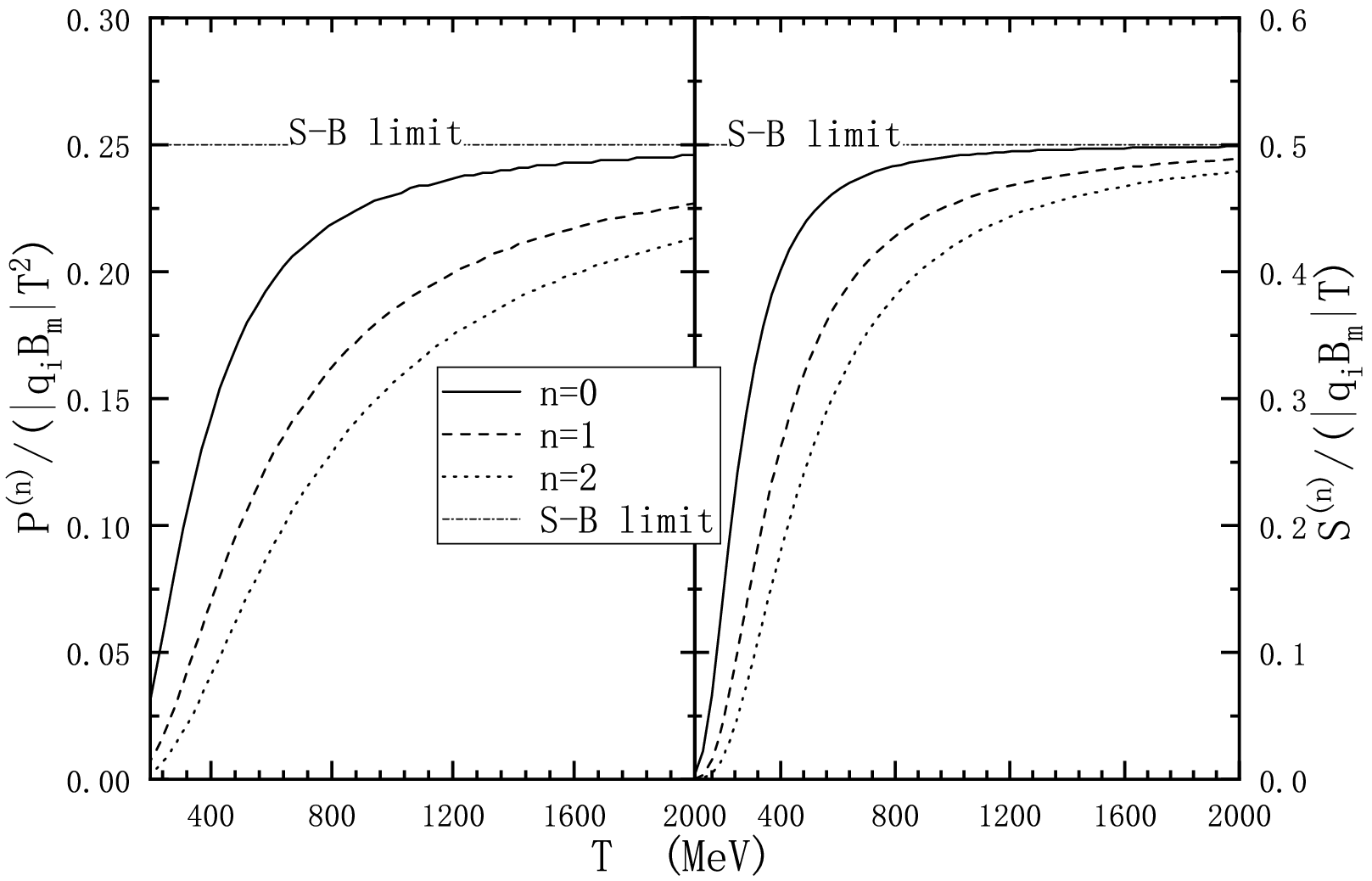} 
\caption{\label{fig:wide}The scaled pressure and entropy of
$i$-flavor quarks lying in Landau levels $n=0$ , 1 and 2. At high
enough temperatures, the three lines approach the Stefan-Boltzman
limit marked by the dash-dotted horizontal line. It is
characteristic that the limit is early approached in the lowest
landau level. }\label{fig:SBlimit}
\end{figure*}

In fact, the S-B limit for the the $n$-th landau level can be
defined as,
\begin{eqnarray}\frac{P^\mathrm{(n)}_\mathrm{SB}}{|q_iB_m|T^2}=\frac{1}{4}.
\end{eqnarray}
In Fig. \ref{fig:SBlimit}, the pressures and entropy of $i$-flavor
quarks in the $n$-th Landau level are shown as the function of
temperatures. At sufficiently high temperatures, the scaled pressure
$P^{(n)}/(|q_iB_m|T^2)$ on the left panel can approach the limit
marked by the dash-dotted horizontal line. In particular, the
pressure from the lowest landau level (n=0) marked by the solid line
is close to the S-B limit line. Moreover, much higher temperature is
required for any excited levels (n=1, 2) to reach the limit. It can
be accounted for by the fact that the higher level leads to the
larger effective mass and therefore results in a larger deviation
from the S-B limit. Correspondingly, the entropy
$S^{(n)}/(|q_iB_m|T)$ in the $n$-level has the S-B limit on right
panel in the strong magnetic field.

\begin{figure*}[htbp]
\includegraphics[width=2.5in]{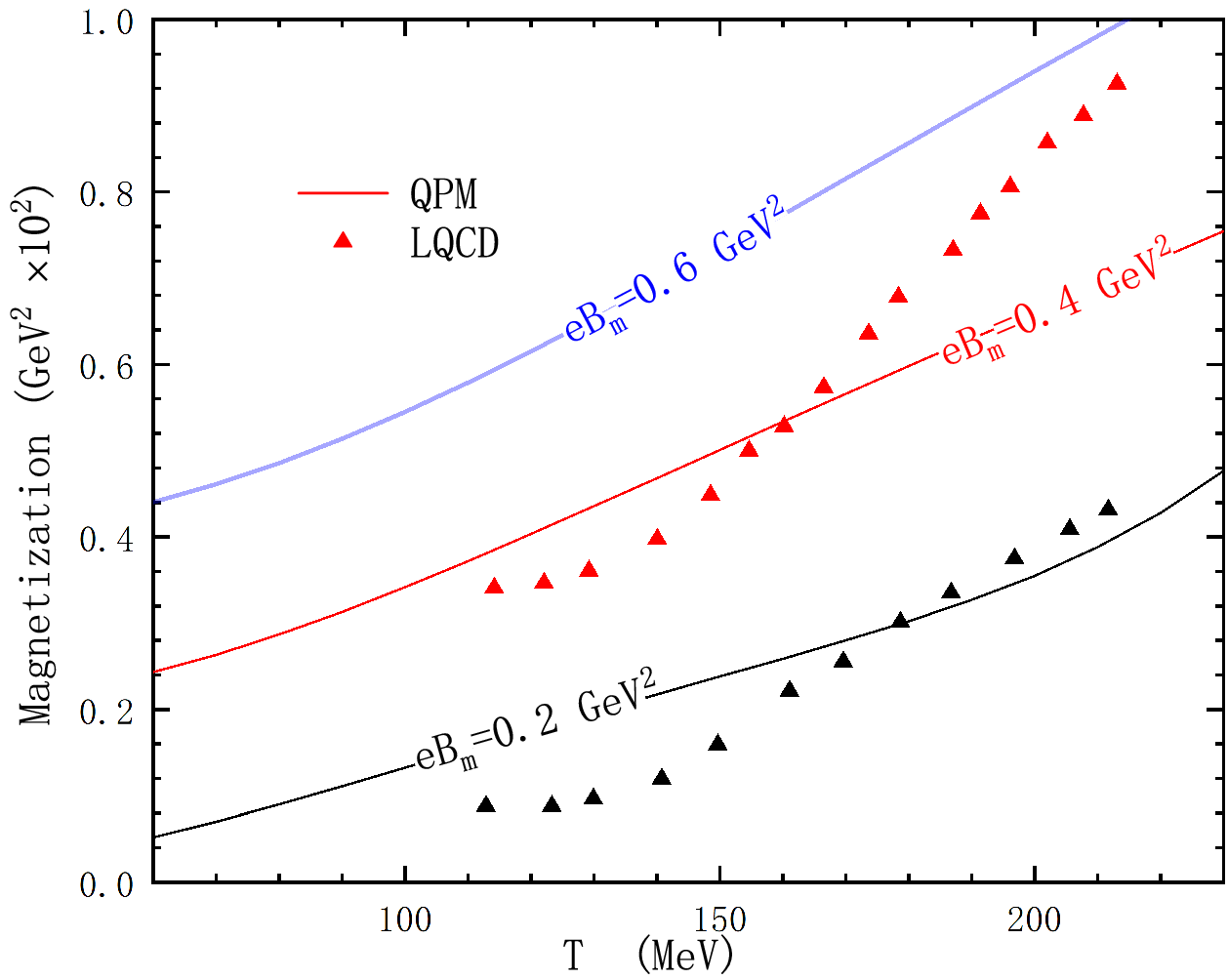}
\caption{\label{fig:wide}The magnetization of the strange quark
matter is shown as a function of the temperature at $eB_m=0.2$, 0.4,
0.6 GeV$^2$. The LQCD result at $eB_m=0.2$ and 0.4 GeV$^2$ is marked
by scattering triangles for comparison
\cite{Bali:2014kia}.}\label{fig:mag}
\end{figure*}

In Fig. \ref{fig:mag}, the magnetization of strange quark matter is
shown as a function of the temperature at fixed magnetic fields
$eB_m=0.2$, 0.4, 0.6 GeV$^2$ marked by the black, red, and blue
solid curves, respectively. The positive value produces the
paramagnetic characteristic for the whole temperature range
\cite{Bali:2013owa}. The magnetization increases with increasing
temperature at fixed magnetic fields. It can be understood that the
more Landau levels at high temperature the stronger the
magnetization.  The effective bag function marked by the red dotted
line enhances the magnetization at finite temperature. The
discrepancy of the magnetization at low temperature is sizable due
to the free magnetic contribution. By comparison with the lattice
result marked by the scattering triangles at $eB_m=0.2$ and 0.4
GeV$^2$ \cite{Bali:2014kia}, it can be concluded that the ascending
trend and the magnetic effect are consistent. In particular,
increase of the magnetic field enhances the magnetization of the
quark matter.

\section{Summary}
In this paper, we have investigated the hot QCD matter exposed to
sufficiently high magnetic fields, which could be generated in RHIC
experiments. The quasiparticle model is extended by including the
free magnetic contribution and the effective thermomagnetic bag
constant, which is self-consistent derived to represent the
confinement. The running coupling constant has been employed to
reflect the asymptotic freedom of QCD. It has been found that the
decrease of the effective bag constant at the high temperature
indicates the occurence of the deconfinement transition. Moreover,
the stronger magnetic field results in a more rapidly decrease of
the effective bag constant, which provides a novel method to account
for the so-called inverse magnetic catalysis effect. Moreover, the
paramagnetic characteristic of QCD is obtained in the quasiparticle
model. The effective bag constant would have an additional
contribution in the new definition of the magnetization due to the
medium effect. It is concluded that the magnetization modified by
the function and free magnetic contributions can only account for
the trend revealed by the lattice result. It would be of some
interest to improve the quasiparticle model to quantitatively
interpret the lattice results in future.

For the quark-gluon plasma in strong magnetic field, the interaction
measure remains larger even at very high temperature and indicates
some interactions are present. Therefore, the usual S-B limit is not
applicable. The deviation form the S-B limit becomes remarkable with
the massive effective mass led by the stronger magnetic field. Not
only the temperature but also the magnetic field are the scale of
energy. We suggested that for single landau level, the S-B limit of
the quark pressure can be defined as $|q_iB_m|T^2/4$. It has been
shown that the lowest landau level is close to the S-B limit, while
higher temperature is required for the excited levels to approach
the limit.

\acknowledgments{ The authors would like to thank support from the
National Natural Science Foundation of China under the Grant Nos.
11875181, 12047571, and 11705163. This work was also sponsored by
the Fund for Shanxi "1331 Project" Key Subjects Construction.}

\end{document}